\documentclass[aps,prd,preprint,showpacs,a4paper,nofootinbib,endfloats*]{revtex4}
\usepackage{latexsym}
\usepackage{graphics}
\usepackage{graphicx}
\usepackage{bm}
\usepackage{amsmath}
\usepackage{amsfonts}
\usepackage{amssymb}

\newcommand{\beq}{\begin{equation}}
\newcommand{\eeq}{\end{equation}}
\newcommand{\beqy}{\begin{eqnarray}}
\newcommand{\eeqy}{\end{eqnarray}}

\newcommand{\slep}{\tilde{\ell}}
\newcommand{\slepbar}{\overline{\tilde{\ell}}}
\newcommand{\smu}{\tilde{\mu}}
\newcommand{\smubar}{\overline{\tilde{\mu}}}

\newcommand{\ee}{e^+e^-}

\newcommand {\gev}{\textrm{ GeV}}
\newcommand{\stauone}{\widetilde{\tau}_1}
\newcommand{\gravitino}{\widetilde{G}}

\begin{document}

\title{Sleptonium at the linear collider and the slepton 
co-next-to-lightest supersymmetric particle 
scenario in gauge mediated symmetry breaking models} 

\author{\sc Nicola Fabiano$^\dag$ and Orlando Panella$^\ddag$}
\affiliation{$^\dag$Dipartimento di Fisica, Universit\`a di Perugia,  
Via Pascoli I-06123, Perugia, Italy
}
 \affiliation{$^\ddag$Istituto Nazionale di Fisica Nucleare, 
Sezione di Perugia, Via A. Pascoli I-06123, Perugia, Italy}

\date{\today}

\begin{abstract}
We discuss the possibility of formation and subsequent detection of a
supersymmetric bound state composed of a slepton--antislepton pair at the next 
linear collider. The Green function method is used within a non-relativistic 
approximation to estimate the 
threshold production cross-section of the $2P$ bound state.  
The parameter space of Gauge Mediated Symmetry Breaking (GMSB) models allow 
a particular  scenario in which a charged slepton 
($\widetilde{e}_R, \widetilde{\mu}_R$ or $\widetilde{\tau}_1$)  is the NLSP.
Within this scenario 
the produced $2P$ bound-state decays, through a dipole transition, into the 
$1S$ ground-state with branching ratio $\approx 100\% $ emitting a very 
soft ($\approx 1 $ MeV) photon which goes undetected.
The spectroscopy of the $1S$-state shows that it decays into 
two photons with $Br \approx 0.5$ up to $m_{NLSP} \approx 1$ TeV. 
Thus NLSP sleptonium threshold production gives rise  to the signal 
$e^+e^- \to 2P \to 1S + ``soft\,\,  \gamma " \to \gamma \gamma$ 
which when compared with the  standard model
two-photon process  ($ e^+e^- \to \gamma \gamma$ )
has a statistical significance ($SS$=signal/noise) which, at an energy 
offset from threshold of $E=20$ GeV, goes from  $SS=11$ to $SS=2$  
when the mass of the NLSP ranges in 
the interval $[100,200]$ GeV.
\end{abstract}
\maketitle
\section{Introduction}
\label{sec:intro}
Despite the enormous success of the standard model (SM) of particle 
interactions based on the gauge group $SU(3)_C \otimes SU(2)_L \otimes U(1)_Y$ 
a large portion of current research efforts in the field of fundamental 
interactions is devoted to the study of signatures of physics beyond 
the SM. 
Between the possible alternatives to the standard model 
its minimal supersymmetric extension (MSSM) is one of 
the theories which has been extensively studied from both the purely 
theoretical and phenomenological aspects~\cite{haberkane}.

Supersymmetry, is the symmetry which relates  fermions to bosons and, while 
appealing because of its potential of solving the 
hierarchy problem of the standard model, must of course be broken, and 
one of the  major issues in supersymmetric theories is the pattern 
of supersymmetry breaking.  
There exist various possibilities
to break supersymmetry. In mSUGRA models it is assumed that supersymmetry 
is broken at the Planck scale $M_P$ and is transmitted to the low-energy 
sector by gravitational interactions only~\cite{mSUGRA}.
In gauge mediated symmetry breaking (GMSB) models supersymmetry 
is broken at relatively lower energy scales and is mediated by 
gauge interactions~\cite{giudicerattazzi}.    
Other possibilities consist of anomaly mediated symmetry breaking (AMSB).
Here the SUSY-breaking occurs also in a hidden sector but it is 
transmitted to the visible sector via the super-Weyl anomaly~\cite{AMSB}. 

Of course at this stage of experimental and theoretical investigations
it is important to study in detail the phenomenological consequence
of each scenario. Object of this work is the GMSB scenario whose theoretical 
basis  and phenomenology has  already been described in the literature, 
(see~\cite{giudicerattazzi} and references therein). 
In particular we recall that GMSB models are characterized by a 
rather peculiar superparticle
mass spectrum. Indeed in these models the almost
massless goldstino/gravitino, $\widetilde{G}$, is the 
lightest supersymmetric particle (LSP) and in addition to the 
neutralino $\chi^0_1$  next-to-lightest supersymmetric particle
       (NLSP) case, large fractions of the
parameter space offer the possibility of having a {\em charged} NLSP which
can be either the  $\stauone$  or $\slep_R\,  (\ell=e,\mu)$ .

In~\cite{ambrosanio1} the phenomenology 
of the production of a pair of NLSP, within GMSB models, was discussed 
in the context of the Cern LEP2 collider. 
In particular it was found 
that in the stau NLSP or charged slepton NLSP cases typical signatures of
NLSP pair production are $\tau^+\tau^- 
E\negthinspace\negthinspace\negthinspace/$, or $\ell^+\ell^- 
E\negthinspace\negthinspace\negthinspace/$ where the missing energy, 
$E\negthinspace\negthinspace\negthinspace/$, is due to the decay of the NLSP
to the almost massless gravitino, $\widetilde{G}$, (LSP).

The object of this study is to propose a novel type of signature 
for the GMSB slepton NLSP scenario, 
assuming that the next LC will operate (at least initially) at energies 
approximatively in the neighborhood of the threshold of NLSP pairs. 
In this case if  $\slep_R$ or
$\stauone$ are the NLSP then the threshold production of sleptonium, 
a bound state of a slepton-antislepton pair, (smuonium or stauonium) 
in $e^+e^-$ collisions should be considered. 
It turns out that such bound state can give rise to the interesting
signature of two photons and {\em practically no missing energy}. 
This signature has not been considered previously. 
Indeed the produced $2P$ state decays with branching 
ratio $\approx 100 \% $ to the $1S$ state
and a photon whose energy is related to the difference of the energy levels
$E_{2P} - E_{1S}\approx 1 $ MeV.
In turn the $1S$ state decays mainly to two photons:
\begin{eqnarray}
e^+e^- &\to& 2P \to 1S + \text{soft\ } \gamma \nonumber\\
&\phantom{\to}& \phantom{2P \to } \, \,\,  \buildrel | \over \!\!\rightarrow  \gamma \gamma
\end{eqnarray}
The threshold production cross section of the 
$2P$ state in $e^+e^-$ collisions is computed using the Green function method 
within a  non relativistic approximation.  The 
decay widths of all  open channels of  the $2P$ 
and $1S$ states are also provided. 
It is concluded that the two-photon signature, when compared with the 
SM process $e^+e^- \to \gamma \gamma$, has a statistical 
significance ($SS = Signal / Noise$) which ranges in the interval 
from $11 - 3$ when the mass of the slepton ($m_{NLSP}$) 
is between $100$ and $300$ GeV. 

The plan of the paper is as follows: 
section \ref{sec:gmsb} provides a review of the GMSB 
supersymmetric model; in section \ref{sec:formscen} 
is discussed the criterion for 
formation of the supersymmetric bound state (sleptonium); section 
\ref{sec:decwidannmode} presents details of the decay channels of the $2P$ 
and $1S$
bound states; section \ref{sec:prodxsect} contains a description 
of the Green function method to estimate the threshold cross-section 
for the production of the $2P$ bound state; 
section \ref{sec:results} presents a discussion of the
statistical significance of the two-photon signature; finally 
section \ref{sec:conclu} presents the conclusions.

\section{GMSB models}
\label{sec:gmsb}

GMSB models of symmetry breaking are perhaps the most promising 
alternative to the SUGRA scheme where SUSY breaking takes place at the 
Planck mass and is then communicated to the low energy sector by 
gravitational interactions. In GMSB  models supersymmetry breaking occurs
at relatively lower energy scales and it is mediated by gauge 
interactions~\cite{giudicerattazzi,martin}.
One nice feature of these models is the automatic suppression of the SUSY
contribution to FCNC and CP-violating processes.

In the simplest version such models are characterized by the introduction
of {\em messenger} chiral superfield which contain 
quarks $\psi_q,\psi_{\bar{q}}$, leptons 
$\psi_\ell,\psi_{\bar{\ell}}$, scalar quarks $q, \bar{q}$ 
and scalar leptons $ \ell,  \bar{\ell}$ (messenger fields). 
All these particles must acquire very large masses as they have not 
been discovered. 
They do so by coupling to a gauge singlet chiral supermultiplet $X$ via the 
superpotential:
\beq
W_{mess} = \lambda_2 X \ell \bar{\ell} + \lambda_3 X q\bar{q}
\eeq
Then one assumes that the scalar component of $X$ and its auxiliary F-term 
acquire a vacu\-um expectation value (VEV) respectively denoted 
$\langle S\rangle$ and 
$\langle F_S\rangle$.  Assuming for simplicity 
degeneracy of the couplings 
($\lambda = \lambda_2=\lambda_3$) and absorbing the coupling $\lambda$  
into  $\langle S\rangle$ and 
$\sqrt{\langle F_S\rangle}$ by defining
$M = |\lambda \langle S\rangle |$ and 
$F=|\lambda \langle F_S \rangle | $ the 
amount of SUSY breaking in the messenger sector
i.e. the mass splitting of the scalar messenger states is found to 
be parametrized as:
\begin{eqnarray}
m^2_{\text{mess. fermions}}&=& 
 M^2 \nonumber\\  
m^2_{\text{mess. scalars}}&=& M^2 \pm F \nonumber \\
\delta m^2_{\text{mess. scalars}} &=&  2 F \qquad \to  \qquad 
\delta m_{\text{mess. scalars}} = \frac{F}{M} =\Lambda \qquad (\, \hbox{if\ } F/M^2 \ll 1) 
\end{eqnarray} 
SUSY breaking is thus apparent in the messenger sector, and is in turn 
communicated to the low energy sector (MSSM sparticles) 
through radiative quantum corrections. Gauginos (and scalar partners) 
acquire their masses, at the messenger scale $M$, 
through one loop (and two loop) Feynman diagrams where virtual 
messenger particles are exchanged~\cite{dimopulos,spmartin,suspect}: 
\begin{eqnarray}
M_G (M) &=& \frac{\alpha_G}{4\pi}\, \Lambda\, g(x)\, \sum_m\, N_R^G(m)  
\qquad G=U(1),SU(2),SU(3)
\label{eq:mass1}\\
{\widetilde{m}}_s^2 (M)& = & 2 \Lambda^2 \, f(x)\,\sum_{G,m} 
\left(\frac{\alpha_G}{4\pi}\right)^2\, C_R^G(s) \, N_R^G(m)
\label{eq:mass2}
\end{eqnarray}
In Eqs.~(\ref{eq:mass1},\ref{eq:mass2}) $x=F/M^2$, $m$ labels the 
messengers and $s$ the MSSM scalar; the functions $f(x)$ and $g(x)$, 
which reduce to $\approx 1$ when $x\to 0$ 
are explicitly given 
in~\cite{ambrosanio1,suspect,spmartin}; $N_R^G$ is the Dynkin 
index  of the gauge representation under which 
the messenger superfields transform, defined by  
$Tr T^aT^b =(N_R^G/2) \delta^{ab} $  the $T^a$ being the generators 
of the gauge group in the representation R\footnote{$N^{U_Y(1)} = (6/5) Y^2$,
 where  $Y=Q_{EM} -T_3$.};
$C_R^G$ are the quadratic Casimir 
invariant  of the same gauge group representation for the MSSM 
scalar field in question and, for the $\bm{N}$
of $SU(N)$), is defined by: 
$\sum_a T^a T^b = C^{SU(N)}_{\bm{N}} \bm{I} = (N^2-1)/2N \bm{I}$
\footnote{$C^{U_Y(1)} = (3/5) Y^2$}. 
$N_R^G$ and $C_R^G$ turn out to be simple algebraic functions of the gauge couplings and the number $n_{\tilde{\ell}}$ and $n_{\tilde{q}}$ of messenger 
fields, see~\cite{suspect,ambrosanio1} for further details.
From Eqs.~(\ref{eq:mass1},\ref{eq:mass2}) one also deduces that 
the scale of SUSY breaking felt in the messenger sector $\Lambda=F/M$ must 
be in the range \[ 10 \text{\ TeV} \le \Lambda \le 100 \text{\ TeV} \] in order
to have sparticle masses in the range of $100$ GeV - 1 TeV. 

A distinctive feature of GMSB models is the fact that the gravitino 
$\gravitino$ may be very light. Indeed $m_{\gravitino}$ is given
by:
\beq
m_{\gravitino}= m_{3/2} = \frac{F_0}{\sqrt{3}M_P'} \approx \left(
\frac{\sqrt{F_0}}{100 \, \hbox{TeV}}\right)^2 \,\, 2.4\, \hbox{eV} 
\eeq  
where $M_P'= (8\pi G_N)^{-1/2} = 2.4 \times 10^{18}$ GeV is the 
reduced Planck mass and $\sqrt{F_0}$ is the fundamental 
scale of super-symmetry breaking (SSB) which  does not coincide with 
${F}$, the scale of SUSY breaking felt by the messenger sector.
The ratio $F/F_0$ depends on how SUSY breaking is communicated to 
the messenger sector. If the communication takes place  via a direct
interaction then $F/F_0$ is given by the corresponding coupling constant which 
by imposing perturbativity arguments  can be shown to 
be smaller than 1~\cite{ambrosanio1}, 
thus giving ${F_0} > F $. If the communication of SUSY breaking 
takes place radiatively  then $F/F_0$ is given by some loop 
factor and thus $F/F_0 \ll 1$. 
It can easily be shown~\cite{ambrosanio2,ambrosanio1} 
that  $F_0$ is only subject to a lower bound which is:
\begin{equation}
\sqrt{F_0}\ge \Lambda
\label{lowerboundonF0}
\end{equation}
which is typically of the order of $10 - 100$ TeV. 

Therefore the  gravitino $\gravitino$ turns out to be the 
lightest super-symmetric particle (LSP). 
In R conserving supersymmetry all sparticles eventually decay to 
the gravitino and in order to compute the decay widths one needs 
the interaction Lagrangian in the gravitino field which can be 
computed in the limit of global supersymmetry ( if $\sqrt{F_0} \ll M_P$) as 
the dominant gravitino interactions come from its spin $1/2$ component 
(the goldstino). It is therefore 
a good approximation to describe the gravitino LSP 
in terms of its spin $1/2$ goldstino component. 
Goldstino interactions contain derivative couplings suppressed by $1/F_0$:
\begin{equation}
{\cal L}_{\text{int}}= - \frac{1}{F_0} \left( 
\bar{\psi_L}\gamma^\mu\gamma^\nu \partial_\nu 
\phi -\frac{i}{4\sqrt{2}}\bar{\lambda}^a\gamma^\mu\sigma^{\nu\rho} 
F^a_{\nu\rho } \right) \partial_\mu\widetilde{G} + h.c.
\label{goldstino-interaction}
\end{equation}
Sparticle phenomenology  (production and decay) is strongly affected by 
the type of the next to lightest super-symmetric particle (NLSP). 
All sparticles will 
decay to a cascade leading to the NLSP which will in turn {\em only } decay to 
the gravitino $\widetilde{G}$  via $1/F_0$ interactions. 
Depending on the values of the parameters, the NLSP can be  either 
the neutralino $\chi^0_1$, the stau $\stauone$, or in restricted regions 
of the parameter space the sneutrino ($\widetilde{\nu}$).\\
Of particular interest in this work is the case of a {\em charged NLSP}. 
In GMSB models the $\stauone$ is always the NLSP but in some 
circumstances it may happen that the mass of the ${\slep_R}$ be closer 
to $m_{\stauone} $ than the mass of the tau ($m_\tau $). If this is the case  
the ${\slep_R}=(\widetilde{e}_R,\widetilde{\mu}_R)$ act effectively 
as a NLSP since the decays 
${\slep_R} \to \ell \stauone^\pm \tau^\mp$ are kinematically forbidden. 
This situation is referred to as slepton co-NLSP scenario, 
and it is more precisely defined  by the condition:
\begin{equation}
\label{co-nlsp}
m_{\slep_R}
< \text{Min}\,[m_{\chi^0_1},m_{\stauone}+m_\tau]\, .
\end{equation}
Within this scenario the sleptonium bound state of a pair 
of $\stauone$ or $\slep_R$  would be
{\em the lightest SUSY state to be produced  in 
a laboratory}. It is therefore interesting to explore throughly all 
possibilities to detect such a bound state at the next linear 
collider (NLC). In the following section we describe the 
spectroscopy in detail.  \\
Within this scenario
the NLSP $ \slep=(\tilde{e}_R,\tilde{\mu}_R, \stauone) $ total width is easily 
determined from Eq.~(\ref{goldstino-interaction}):
\beq
\label{GammaNLSP}
\Gamma_{\slep}=\Gamma(\slep \to \ell \widetilde{G}) = 
\frac{m^5_{\slep}}{16 \pi F_0^2} = 
\left( \frac{ m_{\slep} }{ 100\, \hbox{GeV} } \right)^5 
\left(\frac{100 \, \hbox{TeV}}{\sqrt{F_0}}\right)^4 
\, 2 \times 10^{-3}\, \hbox{eV}
\eeq  
which will be used to establish a criterion for the formation of the 
bound state.

It should be noted that Eqs.(\ref{eq:mass1},\ref{eq:mass2}) are to 
be considered as boundary conditions at the (high) messenger mass scale $M$.
Low energy values of the parameters are to be obtained by running 
renormalization group equations (RGE) down to the electroweak scale. 
This process must of course ensure proper breaking of the electroweak 
(EW) symmetry.
All this is achieved by  the public domain code 
{\tt Suspect}~\cite{suspect}, a software which allows 
for the possibility of choosing
between mSUGRA, GMSB and AMSB models in addition to the general 
unconstrained MSSM. With {\tt Suspect} is possible to perform a scan 
of the parameter space in order to select the scenario that one is 
interested in.

In its simplest version the GMSB model is characterized by a relatively small
number of parameters:
\beq
M,\,\, \Lambda=\frac{F}{M},\,\, n_{\tilde{\ell}}, \,\, n_{\tilde{q}},\,\, \tan(\beta),\,\, sign(\mu)
\eeq 

\begin{figure}[h]
\begin{center}
\includegraphics[angle=0,width=0.9\textwidth]{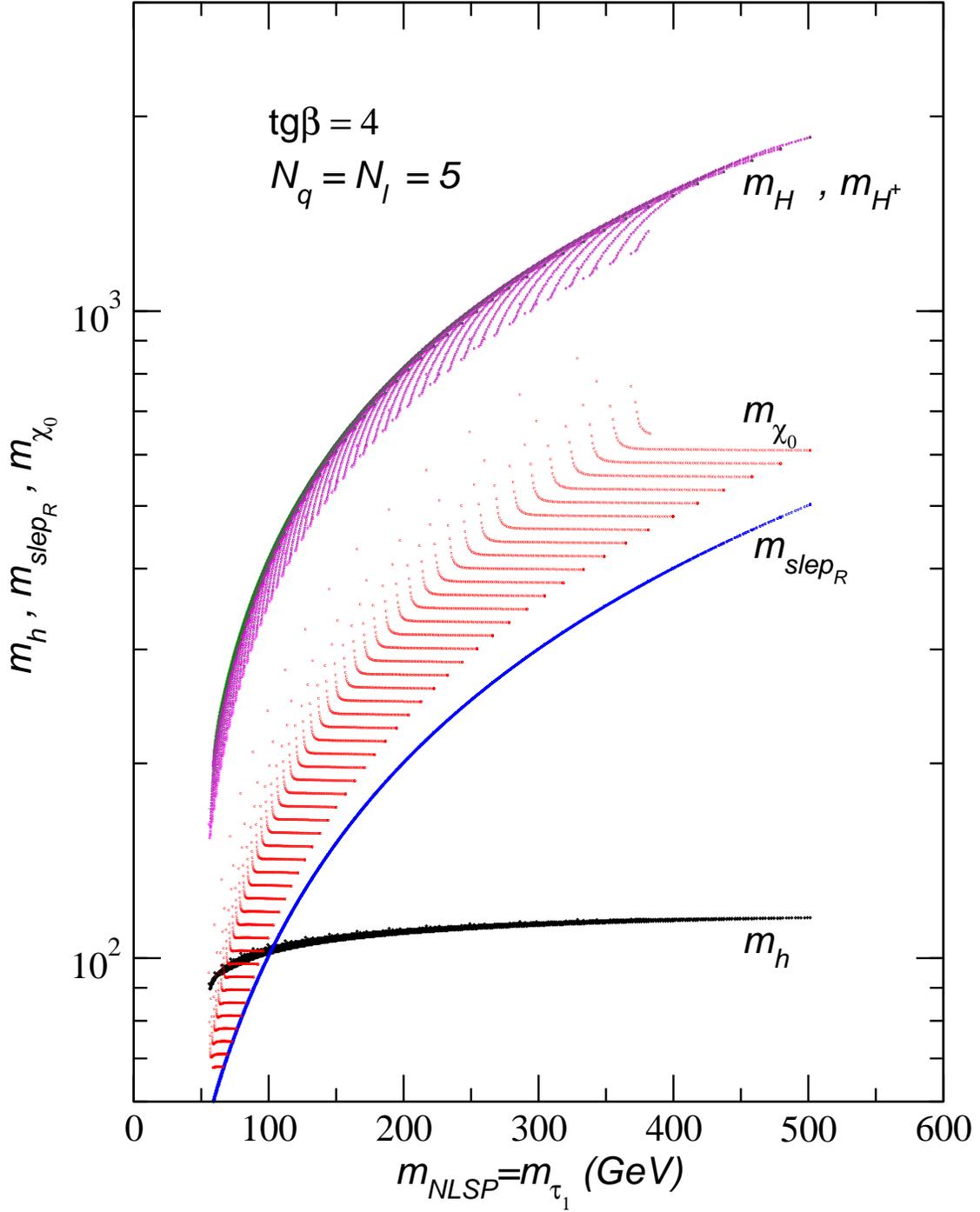}
\end{center}
\caption{Sparticle spectrum in the slepton 
co-NLSP scenario. Only the masses of the 
sparticles which are relevant for our processes are shown 
as a function of the NLSP mass $m_{NLSP}=m_{\tilde{\tau_1}}$. 
What is shown here is a scan of the GMSB parameter space done imposing 
the slepton co-NLSP scenario condition given in Eq.({\protect\ref{co-nlsp}}).}
\label{fig:sparticlespectrum}
\end{figure}

In Fig.~\ref{fig:sparticlespectrum} we show the result of such
a (numerical) study of the parameter space having fixed $\tan\beta=4$ and 
$n_{\tilde{\ell}}= n_{\tilde{q}} = 5$.
The scan has been performed on the plane $M, \Lambda$.
In the figure  
we show the masses of 
$\slep_R (\approx m_{NLSP})$ and the higgs mass states entering into the 
spectroscopy of the bound state decays to be discussed in sec. IV. 
The mass of the neutralino is shown 
for completeness to confirm that it is heavier than $m_{\slep_R}$.

\section{Formation scenario}
\label{sec:formscen}
In this section we will review the creation of the bound state. For the SUSY
case, our assumption will be that the creation of the bound state does not
differ from the Standard Model case, as the relevant interaction 
is again driven by QED and is regulated by the mass of the constituent 
superparticles.
A criterion for the formation of bound states we shall adopt is 
that~\cite{NOI2,ME}
the formation can occur only if the level splitting, which depends upon the
strength of the interaction among the (s)particles, is larger than the natural
width of the would-be bound state. 
It means that the bound state is formed if the following condition is 
satisfied 
\beq
\Delta E_{2P-1S} \ge \widetilde{\Gamma}
\label{eq:critde}
\eeq
where $\Delta E_{2P-1S}=E_{2P}-E_{1S}$, and $\widetilde{\Gamma}$ 
is the width of the would--be sleptonium, which is
twice the width of the single slepton $\widetilde{\Gamma}=2 \Gamma_{\slep} $, 
as each slepton could decay in a fashion
independent from the other. $\widetilde{\Gamma}$ 
is not the total decay width of the sleptonium bound state, 
as it includes only the single smuon decay modes and not the annihilation 
modes. It represents the minimal energy level spread
necessary for bound state formation. If created, the bound state 
will in turn also have its own annihilation decay modes 
(as discussed in sec. IV).\\
For the case of a scalar bound state ({\em sleptonium}), 
we should consider the Coulombic two--body
interaction
\beq
V(r)= - \frac{ \alpha}{r}
\label{eq:coulqed}
\eeq
where the coupling $\alpha$ is the usual fine structure constant of QED.
With this position we are able to compute analytically the energy 
levels and the wavefunctions within a non relativistic approach. From
\beq
E_n = - \frac{m_{\slep}}{4} \; \frac{\alpha^2}{n^2}
\label{eq:nrgy}
\eeq
one infers that 
\beq
\Delta E_{2P-1S} = \frac{3}{16} \alpha^2 m_{\slep}  
\approx  1\, \hbox{MeV}\, \left[\frac{m_{\slep}}{100\, \hbox{GeV}}\right] 
\label{eq:deltae}
\eeq
Contrary to the QCD interaction case~\cite{NOI} the running of the coupling
constant value is not very important, as the relevant scale given by the Bohr
radius, $2/(m_{\slep}\alpha)$, is of $\mathcal{O}(1) \gev^{-1}$. 
The $\Delta E$ value is thus determined only by the mass of 
the slepton. 
This has to be compared to the width of the would-be 
sleptonium $\widetilde{\Gamma} = 2\Gamma_{\slep}$. 
Thus the requirement of formation,  
is obtained inserting Eq.~\ref{eq:deltae} 
and Eq.~\ref{GammaNLSP} in Eq.~\ref{eq:critde}:
\beq
\text{Bound State formation} \iff\sqrt{F_0} \geq 8 \, m_{\slep}
\label{eq:deltaevsf0}
\eeq

We must emphasize that formation criterion adopted here for the sleptonium
bound state is slightly less stringent than the one based
on the revolution time, for which no bound states exist, if the
revolution time, $t_R = 2\pi r/v$, is larger than the lifetime of
the rotating constituents, $\tau = 1/\widetilde{\Gamma}$, 
as shown in~\cite{ME}.

In order to estimate the revolution time, we use the consequences of the virial
theorem, which reads $\langle T \rangle = -\langle V \rangle /2$ for the 
average of kinetic and potential energies respectively. From the expression 
for the energy levels Eq.~\ref{eq:nrgy} we obtain the average speed of the 
constituent slepton, $\langle v^2 \rangle = \alpha^2/(n^2)$, that is
\beq
\langle v \rangle = \frac{\alpha}{ n}
\label{eq:avgspeed}
\eeq
while the average distance of the constituent is given by
\beq
\langle r \rangle = r_B n^2 \left [ 1 + \frac{1}{2} \left ( 1-
\frac{l(l+1)}{n^2} \right )\right ]
\label{eq:avgdist}
\eeq
for a Coulombic potential as in Eq.~\ref{eq:coulqed}. Combining 
Eqs.~\ref{eq:avgspeed} and~\ref{eq:avgdist} we compute the revolution time
for the given state
\beq
t_R = \frac{4  \pi}{m_{\slep}\alpha^2} n^3 \left [ 1 + \frac{1}{2} \left ( 1-
\frac{l(l+1)}{n^2} \right )\right ] \, .
\label{eq:revtime}
\eeq
Thus employing as a formation criterion that the bound state constituents 
life-time $\tau = {1}/{\widetilde{\Gamma_{\slep}}} $ 
be larger than the revolution time leads us (see Eq.~\ref{GammaNLSP}) 
to the inequality:
\begin{equation}
\text{ ($n,l$) Bound state formation} \iff 
\frac{16\pi {F_0}^2}{m_{\slep}^5} 
\ge
\frac{4  \pi}{m_{\slep}\alpha^2} n^3 \left [ 1 + \frac{1}{2} \left ( 1-
\frac{l(l+1)}{n^2} \right )\right ] \, .  
\label{eq:timecomp}
\end{equation}
We obtain two different conditions for the
existence of the $1S$ and $2P$  bound states:
\beq
\text{$1S$ Bound state formation} \iff \sqrt{F_0} \ge 9.15 \, m_{\slep}\, ,
\label{eq:form1s}
\eeq
and
\beq
\text{$2P$ Bound state formation} \iff \sqrt{F_0} \ge 14.7 \, m_{\slep} \, .
\label{eq:form2p}
\eeq
We thus realize that both formation criteria, Eq.~(\ref{eq:deltaevsf0}) and
Eq.~(\ref{eq:form1s},\ref{eq:form2p})
give comparable restrictions on the value of the fundamental scale 
of SUSY breaking $F_0$. 
In order to make sure that the bound state can be produced adopting  
both formation criteria the more conservative criterion of the two 
(that of the revolution time) is chosen. 

For a slepton mass of the order of $100-200$ GeV 
{\em the formation of the bound state(s) is assured} on account of 
Eq.~\ref{lowerboundonF0} ($\sqrt{F_0} \ge \Lambda$) and the 
fact that $\Lambda \approx 10 - 100 $ TeV. 
We stress that the slepton mass $m_{\slep}$ is independent of the energy 
scale $\sqrt{F_0}$ which determines only the strength of the gravitino
interactions.

\section{Decay width and annihilation modes}
\label{sec:decwidannmode}
The scalar bound state formed by a pair of slepton NLSP in an 
$e^+e^-$ collision
will be a $2P$ state with several decay channels. It will  
decay into pair of standard model fermions. 
On the other end being formed by NLSP  it will have only one decay 
channel into super-symmetric particles: 
{\em it will decay (annihilate) into a pair of LSP},  
the almost massless gravitino (goldstino). Finally the 
$2P$ state will decay via dipole interactions to a $1S$ state 
emitting a photon. 
Before discussing  these decay channels in detail we make an 
important remark. Within the slepton co-NLPS scenario the bound state can 
either be $\slep_R^+\slep_R^-$ or $\stauone^+\stauone^-$ as $\slep_R$ 
and $\stauone$ are nearly degenerate in mass. When discussing the 
decay of the bound state however the case of the $\stauone$ is somewhat 
complicated by the fact that left-right 
mixing must be taken into account and  diagrams, which 
are absent in the $\slep_R$ only case,
have to be included. Therefore {\em as a first step we consider only  
a  bound state of} $\slep_R =\widetilde{e}_R,\widetilde{\mu}_R $. 
The case of the $\stauone$ shall be treated on a separate work.\\
In addition while $\widetilde{\mu}$ and $\widetilde{\tau}$ cross 
sections are universal $\widetilde{e}$ pair production suffers 
from destructive interference effects between s-channel and t-channel 
diagrams in the parameter region
relevant for GMSB models (moderate values of $\tan\beta$)~\cite{GFgiudice}. 
For this reason {\em in the following we shall consider only the 
the  bound state of $\widetilde{\mu}_R$} (smuonium).  

The calculation of the partial widths of the decay of the sleptonium 
bound state ($\slep_R^+\slep_R^-$), i.e. $2P \to X$ or $1S \to X$,  is done by 
relating  the amplitude ${\cal M} (B \to X)$ to that of the process 
$\slep_R^+ \slep_R^- \to X$ via a non relativistic model of the bound 
state~\cite{peskin}:
\begin{equation}
\label{nrmodel}
{\cal M} (B \to X) = \sqrt{2M_B} \, \int \frac{d^3\bm{k}}{(2\pi)^3} 
\widetilde{\psi}^*(\bm{k}) \, \frac{1}{2m_{{\slep}_R}} 
\left[ 
{\cal M} \left(\slep_R^+ \slep_R^- \to X\right) 
\right]_{s \to 2m_{{\slep}_R}} 
\end{equation}
$\widetilde{\psi}(\bm{k})$ being the Fourier transform of the  
hydrogen-like wave function of the non relativistic bound state.
The amplitude ${\cal M} \left(\slep_R^+ \slep_R^- \to X\right)$ will in general
be described by one or more tree-level Feynman diagrams an thus its threshold
behaviour $s \to 2m_{{\slep}_R}$ may be inferred. 
Thus for each decay one writes down the amplitudes of the contributing 
Feynman diagrams and then extracts the dependence of the full amplitude 
on $\bm{k}$, the momentum of the constituents sleptons which is assumed to be 
smaller with respect to the mass of the sleptons, $|\bm{k}| \ll m_{\slep_R}$,
giving $\sqrt{s} = 2 \sqrt{m_{\slep_R}^2 +|\bm{k}|^2} \approx 2 m_{\slep_R} \left( 1+|\bm{k}|^2/m_{\slep_R}^2\right) $. 

In $S$-wave decays one obtains an amplitude whose first term in the momentum
expansion is a constant. Then the $\bm{k}$ integration 
in Eq.~(\ref{nrmodel}) gives the wavefunction evaluated at the origin: $\int d^3{\bm{k}}/(2\pi)^3 \widetilde{\psi}^*(\bm{k})= {\psi}^*(0)$.
  In $P$-wave decays one obtains instead an amplitude whose first term 
in the $\bm{k}$ momentum
expansion is linear. Then the $\bm{k}$ integration 
in Eq.~(\ref{nrmodel}) gives the gradient of  
wavefunction evaluated at the origin: 
$\int d^3{\bm{k}}/(2\pi)^3 \bm{k}_i 
\widetilde{\psi}^*(\bm{k}) = \nabla_i {\psi}^*(0)$.

For further details see for example~\cite{JACKSON,drees}. In passing we note
that in the process of comparing the amplitude of the two photon decay mode
of the $1S$ state we find an extra factor of 2 in Eqs. (A1) and (A2) 
of ref.~\cite{drees}, relative to stoponium decay into two 
photons and two gluons, a remark that had already been
made in ref.~\cite{gorbunov}.

\subsection{Decay channels of the $\bm{2P}$ state}

\noindent{\em Decay into gravitinos}: $2P \to \widetilde{G} \widetilde{G}$\\
The process is described by a $t$-channel exchange of a lepton of the 
same flavour of the bound state. The decay width is:
\begin{equation}
\Gamma(2P \to \widetilde{G} \widetilde{G}) = \frac{1}{32\pi^2}\, 
\frac{|R{'}_{2P}(0)|^2}{M_B^4} \, \left(\frac{M_B}{\sqrt{F_0}}\right)^8
\label{eq:decay2grav}
\end{equation} 

\noindent{\em Annihilation into neutrinos}: 
$2P \to \nu_{\ell}\bar{\nu_{\ell}},\,\, (\ell = e,\mu,\tau) $.\\
This is the simplest decay annihilation channel which takes place only 
through the Z boson. In this case there are 
no $t$-channel exchange graphs,  
 even when the flavour of the neutrinos is the same of the sleptonium 
bound state,  since {\em only} left s-leptons ($\widetilde{\ell}_L$) 
do couple charginos and neutrinos. 
The decay width is:
\begin{equation}
\Gamma(2P \to \nu_{\ell}\bar{\nu_{\ell}} ) = \, {\alpha^2}\,
\frac{|R{'}_{2P}(0)|^2}{M_B^4} \, \frac{1}{\cos^4\theta_W}\, f(r_Z,\epsilon_Z)
\label{eq:decay2nu}
\end{equation} 
where $f(x,y)={1}/[{(1-x^2)^2+(xy)^2}]$ and $r_Z=M_Z/M_B; 
\epsilon_Z= \Gamma_Z/M_B$.

\noindent{\em Annihilation into charged standard model 
fermions}
$2P \to f\bar{f}$ \\
First we consider the case that $f$ is either a quark or a lepton 
($\ell' \neq \ell$), $\ell$ being the flavour of the sleptonium bound state. 
When this is the case the decay is through the annihilation into 
$\gamma$ and $Z$-boson only: there are no $t$ channel exchange diagrams.
The decay width is then:
\begin{eqnarray}
\Gamma(2P \to f\bar{f}) &=& 8 \,C_F\,\alpha^2  \frac{|R_{2P}'(0)|^2}{M_B^4}
\sqrt{1-4\frac{m_f^2}{M_B^2}}\left(1 -\frac{m_f^2}{M_B^2}\right)\, \cr
&&\times \left[
Q_f^2 +\frac{c_V^2+c_A^2}{4\cos^4\theta_W}\,f(r_Z,\epsilon_Z) 
+ Q_f\frac{c_V}{\cos^2\theta_W}(1-r_Z^2)\, f(r_Z,\epsilon_Z) \right]
\label{eq:decay2f}
\end{eqnarray}
where: $C_F=3$ for quarks while $C_F=1$ for leptons; $Q_f$ is the 
charge of the fermion in units of $+e$; $c_V$ and $c_A$ are respectively 
the vector and axial coupling of the fermion to the Z-boson: $c_V^f=T^3_f-2Q_f\sin^2\theta_W$ and $c_A^f= T^3_f$;
$m_f$ is the mass of the fermion;
the function $f(x,y)$ is the same that appears in Eq.(\ref{eq:decay2nu}).
One might notice that the above formula reduces to the decay width 
into neutrinos by taking the limit $Q_f=0$ and $m_f=0$.
It also reduces to the formula of refs.\cite{c.nappi,VANROYEN}:
$R_{2P}'(0)$ 
being the derivative of the radial part of the wave function at the 
origin, and $M_B=2m_{\slep}$ is the bound state mass.\\
When the fermion $f$ is the lepton $\ell$ of the same flavour of the slepton 
$\slep_R$ forming the bound state Eq.~(\ref{eq:decay2f}) is to be replaced by the following (the lepton $\ell$ is assumed mass-less):
\begin{eqnarray}
\Gamma(2P \to \ell^+\,{\ell^-}) &=& 8 \alpha^2  \frac{|R_{2P}'(0)|^2}{M_B^4}
\, \left[
1+\frac{1/2+4\sin^4\theta_W -2\sin^2\theta_W}{4\cos^4\theta_W}\,
f(r_Z,\epsilon_Z) \right.\cr
&& \phantom{xxxxxxxxxxx}\left. +\frac{1/2-2\sin^2\theta_W}{\cos^2\theta_W}(1-r_Z^2)
\, f(r_Z,\epsilon_Z) \right.\cr 
&&\phantom{xxxxxxxxxxx} \left. + \frac{G^2}{2} -G +\sin^2\theta_W G \frac{1-r_Z^2}{\cos^2\theta_W}f(r_Z,\epsilon_Z)\,\right]
\label{eq:decay2l}
\end{eqnarray}
where the form factor $G$ describes the diagram of the $t-$channel exchange
of a virtual neutralino. Assuming that the neutralino is mostly bino one has the simplified expression with:
\[G=\frac{4M_B^2}{M_B^2+4m^2_{\chi^0}}\]\\
\noindent{\em Dipole decay  into the ground  state and a photon, 
$2P \to 1S +\gamma$}\\
The decay to the ground state takes place  through a transition  
with emission of a photon. This transition can be computed in the 
long wavelength approximation. 
Indeed the photon momentum $Q=\Delta E_{2P-1S}$, see Eq.~(\ref{eq:deltae}), 
and the bound state dimension (Bohr radius)
$r_B =2/(m_{\slep}\alpha)$ satisfy the relation 
\[
Qr_B = \frac{6}{16}\alpha \approx 2\times 10^{-3} << 1  
\] 
which is the condition that makes the dipole approximation suitable. Then 
a standard quantum mechanics calculation gives~\cite{davydov}:  
\beq
\Gamma(2P \to 1S + \gamma) = \frac{4}{9} \; \alpha  (\Delta E_{2-1})^3
(D_{2,1})^2
\label{eq:decaydipole}
\eeq
where $\Delta E_{2-1}$ is 
the energy of the emitted photon, and
\beq
D_{2,1}= \langle 2P | r | 1S \rangle=
\int_0^{\infty} dr r^3 R_{1S}(r) R_{2P}(r)
\label{eq:dipole}
\eeq
is the dipole moment (see~\cite{NOI} and references therein).
The wave functions to be used are the one of the Coulombic model, given by
\begin{eqnarray}
R_{1S}(r) &=& 2 \left ( \frac{1}{r_B} \right )^{3/2} 
\exp \left (-\frac{r}{r_B} \right )\\
R_{2P}(r)& = &\frac{1}{\sqrt{3}} \left ( \frac{1}{2r_B} \right )^{3/2} 
\frac{r}{r_B} \; \exp \left (-\frac{r}{2r_B} \right )
\label{eq:coulombwf}
\end{eqnarray}
where
$r_B$ is the {\em Bohr radius} defined as 
$r_B=2/(m_{\slep}\alpha)=4/(M_{B}\alpha)$.
Using the above wavefunctions  one obtains the following expressions
for the dipole decay mode:
\begin{eqnarray}
\Gamma(2P \to 1S + \gamma) &=&  \alpha^5 M_B \frac{64}{6561}\, \approx 
\, \alpha^5 M_B \, 10^{-2}
\label{dipoledecay}
\end{eqnarray}
Then we have to compute the total decay width to fermions:
\begin{equation} 
\Gamma({2P \to fermions}) = \sum_{f=q,\ell,\ell'} \Gamma(2P \to f\bar{f}) +
\Gamma(2P \to \widetilde{G} \widetilde{G}) 
\label{eq:gamma2fermions}
\end{equation}
The derivative in the origin of the radial wavefunction of the $2P$ 
state is easily computed and it follows that:
${|R_{2P}^{'}(0)|^2}/{M_B^4} = {\alpha^5 M_B}/{24576} $.
Inserting the results of the Eqs.~(\ref{eq:decay2grav},\ref{eq:decay2nu},\ref{eq:decay2f},\ref{eq:decay2l}) into Eq.~(\ref{eq:gamma2fermions}) 
one obtains:
\begin{equation}
\Gamma(2P\to fermions) = {\alpha^5 M_B} \, k\, 10^{-5}
\end{equation}
where $k$ is a numerical constant of order unity with a mild dependence 
on the mass of the bound state $M_B$, the fundamental scale of SUSY breaking
$\sqrt{F_0}$ and the mass of the neutralino $m_{\chi^0}$.  
It then follows that the
branching ratio of the dipole decay is $ Br(2P\to 1S +\gamma) = 100\%$ 
to within one part in $10^3$.
For all practical purposes the $2P$ state will decay with probability $1$ 
to the ground state $1S$ emitting one photon with an energy of a few 
MeV.

\subsection{Decay channels of the $\bm{1S}$ ground state}

\noindent{\em Decay into two photons.}\\
In this case since the photon is described by transversely polarized 
states, in the non-relativistic limit the {\em t-} and {\em u-channel} 
diagrams do not contribute and only the sea-gull 
diagram survives. The decay width is given by: 
\beq
\Gamma (1S \to \gamma\gamma)=2\, \alpha^2  \frac{|R_{1S}(0)|^2}{M_B^2}
\label{eq:decay1s} 
\eeq

\noindent{\em Decay into $\gamma Z$.}\\
Again the fact that the photon does not have longitudinal polarization states
selects, in the non relativistic limit, only the sea-gull diagram. 
The decay width is given by:
\beq
\Gamma (1S \to \gamma Z)= 4 \, \alpha^2 \, 
\frac{\sin^2\theta_W}{\cos^2\theta_W} \frac{|R_{1S}(0)|^2}{M_B^2}\, 
\left(1-\frac{M_Z^2}{M_B^2}\right)
\label{eq:decay1S_gZ} 
\eeq
\noindent{\em Decay into $Z Z$.}\\
Here there are four diagrams that give non zero contribution. The diagram with 
 a $\slep_R$ in the $t$-channel and its exchange contribute only for the 
longitudinal polarization states of the Z gauge boson. 
The sea-gull term ($\tilde{\ell}_R\, \tilde{\ell}_R - ZZ$) and the 
$s$-channel higgs exchange give non zero 
contribution for all type of the Z boson polarization. The decay width is 
found to be: 
\begin{eqnarray}
\Gamma({1S\to ZZ}) &=& \frac{\alpha^2}{2} \,\tan^4\theta_W \, 
\frac{|R_{1S}(0)|^2}{M_B^2}\, \beta_Z \left\{ |F+2|^2 
\left(3-\frac{M_B^2}{M_Z^2} +\frac{1}{4}\frac{M_B^4}{M_Z^4}\right) 
\right.\nonumber\\ 
&&\phantom{xxxx} \left. +G^2\left(\frac{1}{4}\frac{M_B^2}{M_Z^2} -1\right)^2
-2G \left(\Re e{F} +2\right) \left[1-\frac{3}{4}\frac{M_B^2}{M_Z^2}+\frac{1}{8}\frac{M_B^4}{M_Z^4}\right] \right\}
\end{eqnarray}
where:
\begin{eqnarray*}
\beta_Z&=& \sqrt{1-4\frac{M_Z^2}{M_B^2}}\\
F&=& \sum_{i=1,2}\, c_i \,
\frac{M_Z^2}{M^2_B -M_{H_i}^2+i\Gamma_{H_i^0}M_{H_i^0}}\\
G&=&4 M_B^2/(M_B^2-2M_Z^2)\\
c_1&=&\cos(\beta-\alpha)\cos(\beta+\alpha)/\sin^2\theta_W\\
c_2&=&-\sin(\beta-\alpha)\sin(\beta+\alpha)/\sin^2\theta_W\\
\end{eqnarray*}
while $F$ is a form factor arising from the diagrams with $s$-channel Higgs exchange, $G$ arises from the $t$-channel $\slep_R$-exchange diagrams.\\ 
\noindent{\em Decay into $W^+W^-$.}\\
In this case the $t$-channel exchange of a sneutrino is absent (no 
$\widetilde{\nu}_R$) as well as the sea-gull term 
($\tilde{\ell}_R\, \tilde{\ell}_R - WW$), and only the $s$-channel Higgs 
contribution is present. The partial decay width is:
\begin{equation}
\Gamma({1S\to W^+W^-}) = \frac{\alpha^2}{2} \, \frac{|R_{1S}(0)|^2}{M_B^2}\, |F'|^2\, 
\beta_W\, 
\left[ 3 -\frac{M_B^2}{M_W^2} + 
\frac{1}{4}\frac{M_B^4}{M_W^4} \right]
\,
\end{equation}
where:
\begin{eqnarray*}
\beta_W&=& \sqrt{1-4\frac{M_W^2}{M_B^2}}\\
F'&=& \sum_{i=1,2}\, c_i \,
\frac{M_Z^2}{M^2_B -M_{H_i^0}^2+i\Gamma_{H_i^0}M_{H_i^0}}\\
c_1&=&\cos(\beta-\alpha)\cos(\beta+\alpha)\\
c_2&=&-\sin(\beta-\alpha)\sin(\beta+\alpha)
\end{eqnarray*}
\noindent{\em Decay into $hh$.}\\
Within the minimal supersymmetric version of the higgs sector there are 
five higgs states: three neutrals, ($h, H, A$) and two charged, ($H^{\pm}$).
Here we consider only the decay of the $1S$ state into a pair of the 
lightest Higgs states ($h$).
The process receives contribution from three diagrams: 
a) $t-$channel $\slep_R$-exchange; 
b) $s$-channel higgs exchange $(h,H)$ [CP invariance forbids $s$-channel 
exchange of $A$]; 
c) sea-gull term $\slep_R\slep_R-hh$. 
There are no diagrams with $s$-channel exchange of $Z$
 since Bose symmetry forbids $ZH_i^0H_i^0$ couplings.
The decay width is found to be:   
\begin{equation}
\Gamma({1S\to hh }) = \frac{\alpha^2}{2\cos\theta_W^4} \, 
\frac{|R_{1S}(0)|^2}{M_B^2}\, 
\beta_h\, \left|Y\right|^2
\end{equation} 
\begin{eqnarray*}
\beta_h&=& \sqrt{1-4\frac{M_h^2}{M_B^2}}\\
Y&=&\sum_{i=1,2} c_i \frac{M_Z^2}{M_B^2-M_{H_i^0}^2+i\Gamma_{H_i^0}M_{H_i^0}}
-\frac{\cos(2\alpha)}{2} +\sin^2\theta_W\sin^2(\alpha+\beta)
\frac{4M_Z^2}{2M_h^2-M_B^2}\\
c_2&=&(3/2)\sin^2(\beta+\alpha)\cos(2\alpha)\\
c_1&=&2\sin(2\alpha)\sin(\alpha+\beta)\cos(\alpha+\beta)-\cos(2\alpha)
\cos^2(\alpha+\beta)
\end{eqnarray*}

In Fig.~\ref{fig:partialwidths} the partial widths of the various decay 
channels of the $1S$ state are shown with respect to the bound state mass 
$M_B$. We see that the two photon channel always dominates. 
The branching ratio $Br(1S\to \gamma\gamma)$ is $Br=0.65$ at 
$m_{\slep_R} = 100 $ GeV and  $Br\approx 0.52$ up to $m_{\slep_R} = 1000 $ GeV,
 as shown in Fig.~\ref{fig:branching}.

The detection of the $P$ wave bound state is therefore associated to the
emission of a soft photon plus a subsequent emission of two hard photons given
by the decay of the scalar ground state.
\section{cross section}
\label{sec:prodxsect}
\subsection{Slepton pair production cross-section}
We use the notation $\alpha_{\slep}= (3\tan \theta_W-\cot \theta_W)/4$
and $\beta_{\slep}= (\tan \theta_W+\cot \theta_W)/4$.
The production cross section for scalar sleptons - except for
selectrons - is given by~\cite{TATA}
\beq
\sigma(\ee \to \slep_i^+ \slep_i^-)= \frac{\pi \alpha^2}{12}\beta^3
\left [ \frac{4}{s}  + 
\frac{A_{\slep_i}^2(\alpha_{\slep}^2+\beta_{\slep}^2)s-4\alpha_{\slep}
A_{\slep_i}(s-M_Z^2)}
{(s-M_Z^2)^2+(M_Z\Gamma_Z)^2} \right]
\label{eqsigmatoslantisl}
\eeq
with $i=L,R$; $A_{\slep_L} = 2(\alpha_{\slep}-\beta_{\slep})$ or 
$A_{\slep_R} = 2(\alpha_{\slep}+\beta_{\slep})$ for left- and right- handed
sleptons respectively.

It is useful to write down this expression of the $P$ 
wave cross section in the following manner:
\beq
\sigma(\ee \to \slep \slepbar) = \frac{4 \alpha^2}{3s} \;\beta^3 \; d(s)
\label{eq:sigmabschem}
\eeq

\subsection{Bound state production cross-section}
We shall write the threshold cross section of the bound state in terms
of its Schr\"odinger Green function. To review briefly the method, explained in
more detail in~\cite{ME2}, consider the bound state described by a
Schr\"odinger equation with a suitable potential $V(\bm{x})$. The threshold
cross section is then proportional to the imaginary part of
derivative taken at the origin of 
the $P$ wave  Green function of the problem 
$G_1(\bm{x},\bm{y},E)$. $E$ is the energy displacement from
threshold, and the finite width of the state is taken into account by the
substitution $E \to E +i \Gamma$.

The cross section for the production of a bound state can be  normalized to the QED process $e^+e^- \to \mu^+ \mu^-$:  
\begin{eqnarray}
R &=&  \frac{\sigma (e^+e^- \to 2P)}{\sigma(e^+e^- \to \mu^+ \mu^-)} \\
&=& \frac{4\pi}{m^4_{\slep}} \;d(s) \;
 {\Im}m \left [ {\sf Tr} \left . \frac{\partial}{\partial {x_i}}\, 
\frac{\partial}{\partial {y}_j}\,  G^{(1)}(\bm{x},\bm{y},E)\,  \right ] 
\right |_{\bm{x}=0,\bm{y}=0}
\label{eq:R}
\end{eqnarray}
where $d(s)$ is the usual expression of the Born cross section for the process 
$e^+e^- \to \smu \smubar$ written in Eq.~(\ref{eqsigmatoslantisl}).
In our investigation the interaction among the two superpartners is 
driven by a Coulombic interaction with
\beq
V(r) = -\frac{\alpha}{r}
\label{eq:coulomb}
\eeq
Adapting the notations of~\cite{PP} we set $E = \sqrt{s}-2m_{\smu}$ as the
energy displacement from threshold, $k^2=-m_{\smu}E$, the wavelength 
$\lambda=9m\alpha/8$, and the wave number $\nu=\lambda/k$.
Now the explicit form for the $\ell=1$ Green function $G_1$ 
of Eq.~(\ref{eq:R}) takes the form

\begin{equation}
G_1(0,0,k)
= \frac{m_{\smu}}{36 \pi} \lambda \left \{ 2(k^2 - \lambda^2) \left [
\frac{k}{2 \lambda}  + \ln \left ( \frac{k}{\mu_f}\right ) + 2 \gamma_E -
\frac{11}{6} + \psi_1(1-\nu)\right ] + \frac{k^2}{2}\right \} 
\label{eq:g1}
\end{equation}
$\gamma_E$ is Euler's constant, $\psi_1$ is the digamma function, 
logarithmic derivative of the $\Gamma$ function and $\mu_f$ is a soft scale
estimated from a relativistic framework (see~\cite{ME2} and references therein).
The derivative of Eq.~(\ref{eq:g1}) needed for computing the cross 
section Eq.~(\ref{eq:R}) has a simple expression, as we have
\beq
{\sf Tr}\, \left .\frac{\partial}{\partial {x}_i}\, \frac{\partial}{\partial 
{y}_j}\, G^{(1)}(\bm{x},\bm{y},k) \right |_{\bm{x}=0,\bm{y}=0} = 
9G^{(1)}(0,0,k)\,\, .
\label{g1der}
\eeq
From Eq.~(\ref{eq:g1}) one can readily notice that the 
leading term in the cross
section Eq.~(\ref{eq:R}) for large $E$ is given by $k^3$, whereas the peaks
of the bound state energy levels are determined by the digamma function 
in $\nu$.

The Green function method is a non-relativistic procedure. We have therefore to
ensure that the velocity of the sleptonium constituents is low enough in order to keep
relativistic corrections negligible. Starting from the parametrization of the
center of mass energy $\sqrt{s} = 2 m_{\slep} +E$ and assuming an upper value
for the constituent velocity $\beta_{MAX}$ we obtain an upper bound for
the energy offset from threshold
\beq
E \le E_{MAX} = m_{\slep} \beta_{MAX}^2
\label{eq:betamax}
\eeq
by means of a series expansion in $E$. This relation translates to the maximal 
allowed value of Lorentz boost parameter $\gamma$
\beq
\frac{E_{MAX}}{m_{\slep}} = \frac{\gamma^2_{MAX}-1}{\gamma^2_{MAX}}.
\label{eq:gammamax}
\eeq
We thus define the non-relativistic domain  by imposing that the  value 
of $\gamma$ differs from $1$ by less than $\approx 10$\%. This then gives $\beta_{MAX}\approx 1/\sqrt{5}$ and therefore, for 
a slepton mass $m_{\slep}= 100$ GeV, an energy offset
from threshold equal to $E_{MAX}=20$ GeV.
From eq.~\ref{eq:betamax} one could also observe that the acceptable 
(non-relativistic) threshold energy range increases when larger 
values  of the constituent mass are considered.

In addition we emphasize that the entire procedure of treating the 
bound state breaks down away from threshold, independently of where 
relativistic corrections may become important.

\section{Results and discussion}
\label{sec:results}

We shall analyse the threshold behaviour of cross section 
for a range of masses
and SUSY parameters for which the bound state formation is envisaged, as
discussed in section~\ref{sec:decwidannmode}. Following~\cite{ME2} we observe
that the relevant region for our analysis is the one above threshold, i.e. for
$E>0$. In fact the region below threshold, $E<0$, is characterised by
peaks in the cross section located at the discrete energy values of bound 
states. Their width is given by the annihilation modes which, as shown 
in~\ref{sec:decwidannmode} is of the order of the eV at most. From
eq.~(\ref{eq:nrgy}) one can estimate the separation of the discrete peaks. They
merge when the peaks are not distant enough
\beq
\frac{m}{4} \alpha^2 \left [ \frac{1}{n^2} -\frac{1}{(n+1)^2} \right ]
\sim \Gamma
\label{eq:mergepeaks1}
\eeq
The last resolved peak has a quantum number $n$ given by
\beq
\frac{2n +1}{n^2(n^2+1)} \sim \frac{4\Gamma}{m \alpha^2} \ll 1
\label{eq:lastpeak}
\eeq
Due to the beam energy spread of the collider, much larger than a few 
eV of the natural width of the state and of the order of 
few GeV~\cite{TESLA} this structure cannot be observed. The difference from the
usual Born cross section results therefore should be sought for $E>0$.

The effect of the bound state to the cross section at threshold is to accumulate
and merge the peaks towards the $E=0$ value, giving a larger result than the
naive Born cross section, as we could see in 
Figs.~\ref{fig:greenvsbornR} and~\ref{fig:greenvsbornR200}.\\

\subsection{The signal and its statistical significance}
Let us now discuss in some detail the signal 
\begin{equation}
e^+e^- \to 2P 
\to 1S + \text{``soft $\gamma$''} \to \gamma \gamma
\end{equation} 
where the soft photon is assumed to be undetected since its energy is of 
only 1 or 2 MeV, see Eq.~\ref{eq:deltae}. 
Indeed it is known~\cite{tesla2} 
that the photon energy resolution of the calorimetric detector
is, for low-energy photons, of the type 
 $\delta E/E \approx 0.11/ \sqrt{E/\text{GeV}}$ which implies 
that $\delta E/E \approx 1 $ for $ E=10$ MeV. 
Therefore the 1-2 MeV soft photon of our signal
will surely be much below the energy resolution and will be undetected.
 
The observed final state is therefore {\em two hard back-to-back 
photons}.
This two photon signal of the production of the $2P$ sleptonium 
bound state is to be compared with the QED two photon 
process $e^+e^- \to \gamma \gamma$ which is expected to be the 
dominant background.
The number of events of the signal can thus be estimated by:
\begin{equation}
N_{sign.} =  {\cal L} \sigma_{B} \, Br (2P \to 1S +\gamma) \, 
Br (1S \to \gamma\gamma) 
\end{equation}
where $\sigma_B$ is the threshold cross section for the 
production of the bound state computed with the Green function method.
Given the fact that $Br (2P \to 1S +\gamma) \approx 1$ we have 
\begin{equation}
N_{sign.} =  {\cal L} \sigma_{B} \, 
Br (1S \to \gamma\gamma)  
\end{equation}
The statistical significance $SS$ of the signal can thus be estimated by:
\begin{equation}
SS = \frac{N_{signal}} {\sqrt{N_{background}}}= \frac{{\cal L} \sigma_{B} \, 
Br (1S \to \gamma\gamma)} {\sqrt{{\cal L}\sigma_{\gamma\gamma} } }
\end{equation} 
A comment is in order at this point. The two-photon decay of the bound state
is isotropic in the rest frame of the bound state which however will 
approximatively coincide with the laboratory frame as we are assuming 
threshold production with $E\le 20$ GeV. The distortion introduced in 
the laboratory system 
when boosting the isotropic distribution of the two-photon signal from the 
bound state rest frame depends on the 
$\beta(M)$ of the bound state.
Indeed in the bound state rest frame let $\theta^*$ be the angle of the direction
of the two outgoing (back-to-back) photons:
\begin{equation}
\frac{d\sigma_B}{d\cos\theta^*}= \frac{\sigma_B(s)}{4}
\end{equation} 
where the phase space integration has been reduced of a factor of $1/2$ due to the identical particles in the final state.
The distribution can be boosted~\cite{landau} to the laboratory frame where the bound state
has a velocity $\beta(s), \sqrt{s}=2 m_{\slep} + E$, $E$ 
being the offset from the 
threshold:
\begin{equation}
\frac{d\sigma_B}{d\cos\theta}= \frac{\sigma_B(s)}{4}\frac{1-\beta^2}{(1-\beta\cos\theta)^2}
\end{equation}
where now $\theta$ is the direction of the photon in the laboratory frame.\\
On the other end the angular differential distribution of the photons for the QED process $e^+e^- \to \gamma \gamma$ is given by:
\begin{equation}
\frac{d\sigma_{\gamma\gamma}}{d\cos\theta} = \frac{\pi\alpha^2}{s}\frac{1+\cos^2\theta}{1-\cos^2\theta}
\end{equation}
which is symmetric and peaked in the forward and backward directions. Again a factor of $1/2$ has been included in the phase-space factor due to the identical particles of the final state.
Applying an angular cut  in the forward and backward directions 
\[\theta_0< \theta< \pi-\theta_0\] it is possible to suppress the background in 
such a way as to obtain an interesting statistical significance. 
We define a cut dependent statistical significance:
\begin{equation}
SS(\theta_0) = \frac{N_{sig}(\theta_0)}{\sqrt{N_{back}(\theta_0)}} =
\frac{{\cal L}\times Br(1S\to \gamma \gamma) \int_{\theta_0}^{\pi-\theta_0} 
\, \sin\theta d\theta \, (d\sigma_B/d\cos\theta)}{\sqrt{{\cal L}
\int_{\theta_0}^{\pi-\theta_0}\, \sin\theta d\theta 
\,(d\sigma_{\gamma\gamma}/d\cos\theta)}}
\end{equation}
Defining $z_0=\cos\theta_0$ one easily finds:
\begin{equation}
SS(z_0) =\frac{\sqrt{\cal L} \sigma_B(s)}{\sqrt{2\pi\alpha^2 /s}} \times 
Br(1S\to \gamma \gamma) \times \frac{\frac{1-\beta^2}{1-\beta^2z_0^2}\frac{z_0}{2}}{\sqrt{\log\left(\frac{1+z_0}{1-z_0}\right)-z_0}}
\end{equation}
The statistical significance as  function of the angular cut is shown in 
Figs.~\ref{fig:greenvsnoiseR} and~\ref{fig:greenvsnoiseR200}. 
We observe  a very mild dependence on 
$\theta_0$ apart from values of $\theta_0 \approx 0$ 
and $\theta_0 \approx \pi/2 $. For these limiting values the statistical 
significance approaches zero. When $\theta_0 \to 0$, $  SS(\theta_0) \to 0 $ 
since  the QED two photon background total cross-section is divergent. On 
the other hand when $\theta_0 \to \pi/2$ one is integrating over a 
region of phase space of vanishing measure so that both cross sections
vanish as $z_0=\cos\theta_0$ in this limit.
One notices that for values of $\theta_0$ in the range 
$0.4 \le \theta_0 \le 0.8$   the statistical significance is almost constant 
and ranges from $SS\approx 4$ to $SS\approx 11$ for values of the energy 
offset respectively of $E=10$ GeV and $E=20$ GeV, and for a fixed slepton mass
$m_{\slep_R} =100$ GeV.

In Fig.~\ref{fig6} we show the dependence of the statistical significance
$SS(\theta_0)$ computed with $\theta_0= 0.55$ which corresponds to the maximum
of the curves in Fig.~\ref{fig:greenvsnoiseR} for two given values of the energy offset $E=10$ GeV and $E=20$ GeV as function of the sleptonium 
bound state $M_B=2 m_{\slep_R}$.
We see that even going to higher values of the slepton mass 
$m_{\slep} \approx 200$ GeV the statistical significance is still at 
an interesting value $SS\approx 2$ (at $E=20 $ GeV). 

It should also be checked that along with interesting values of 
the statistical significance one has also interesting absolute values of the 
cross sections, i.e. an observable number of events.
Indeed from Fig.~\ref{fig:greenvsbornR} at $E=20$ GeV, $m_{\slep_R}=100$ GeV, 
one can see that $\sigma_B \times BR(1S\to \gamma\gamma)  \approx 80  \times 0.63 $ fb $\approx 50$ fb. This would correspond to a total number of signal events 
$N_{sig} \approx 5000 $ assuming an annual integrated luminosity 
$L_0 = 100 $ fb$^{-1}$.   
For $E=20$ GeV, $m_{\slep_R}=200$ GeV one finds  
$\sigma_B \times BR(1S\to \gamma\gamma) 
 \approx 8  \times 0.63 $ fb $\approx 5$ fb. This would correspond to a total 
 number of signal events $N_{sig} \approx 500 $,  ten times lower 
 than the $100 $ GeV mass case.
The above estimates have been done using the total cross section $\sigma_B$. 
We have checked that taking into account 
the distortion of the boost to the laboratory system from the rest 
frame of the bound state, where the two-photons are isotropically distributed,
changes the above estimates by an amount which is at most of $1 \%$.   

\begin{figure}[h]
\begin{center}
\includegraphics[angle=0,width=0.9\textwidth]{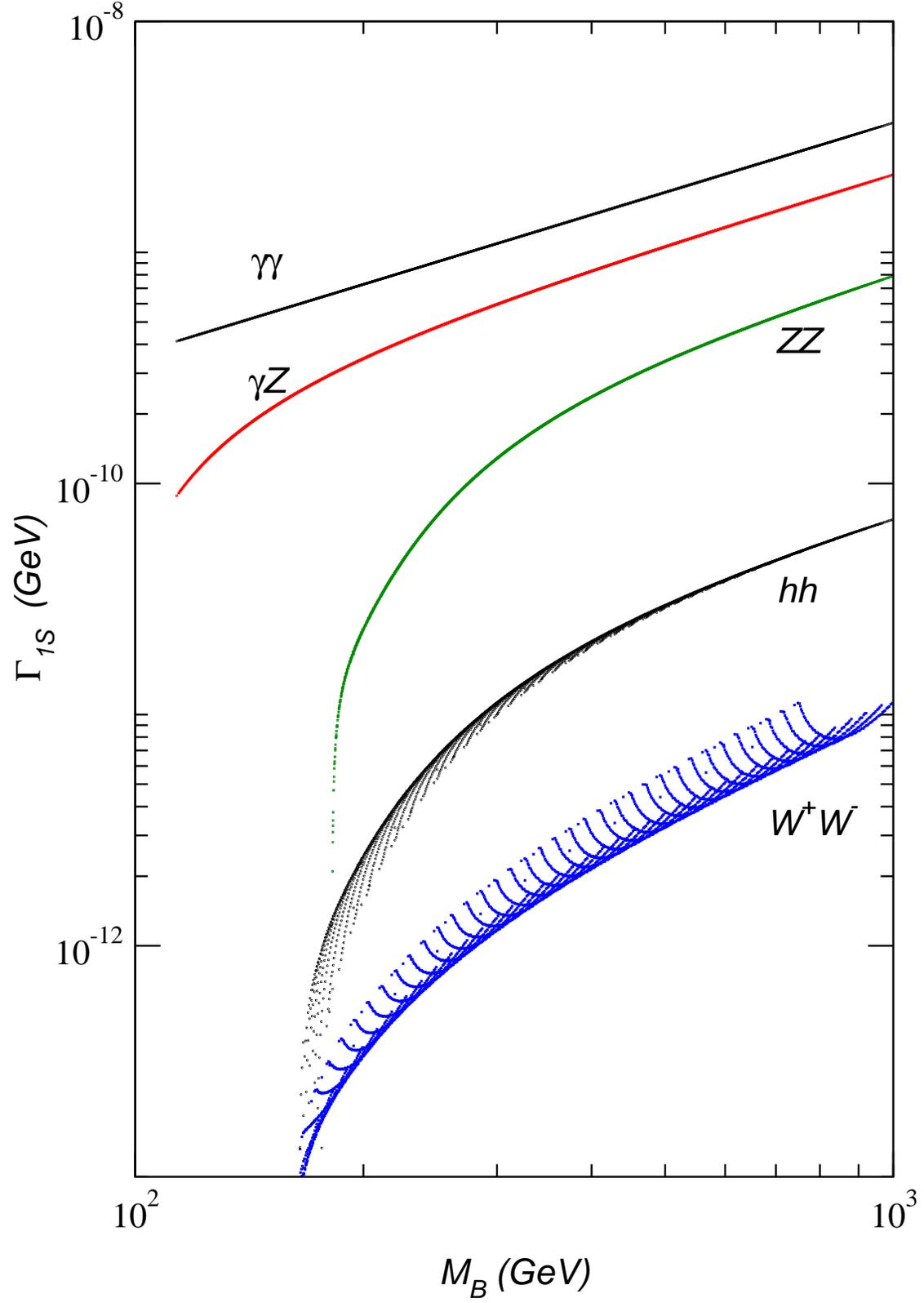}
\end{center}
\caption{Partial widths of the various decay channels of the $1S$ state,
as a function of the bound state mass $M_B=2m_{NLSP}$.}
\label{fig:partialwidths}
\end{figure}

\begin{figure}[h]
\begin{center}
\includegraphics[angle=0,width=0.9\textwidth]{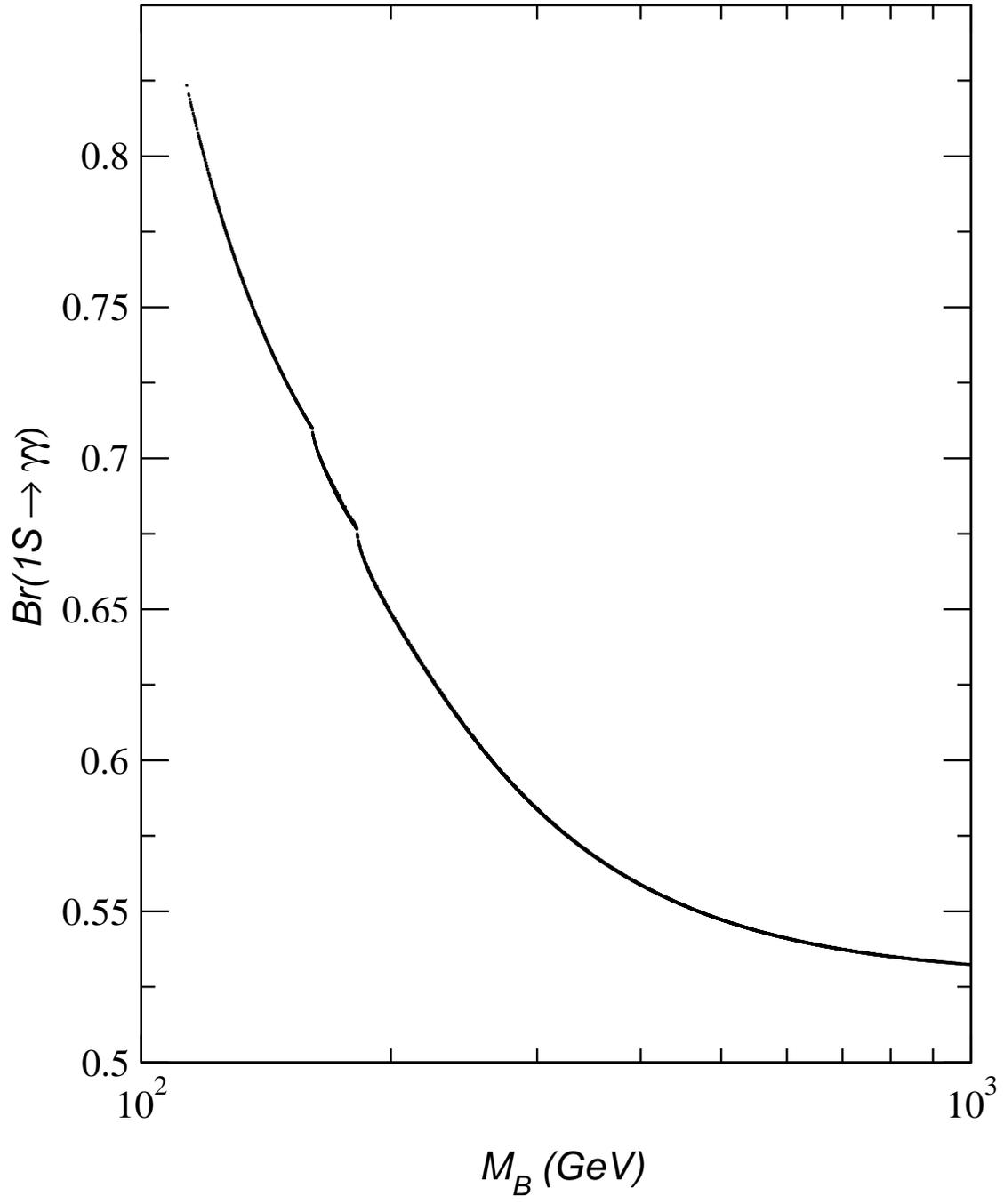}
\end{center}
\caption{Branching ratio of the $1S \to \gamma \gamma$ decay channel 
as a function of the bound state mass $M_B=2m_{NLSP}$.}
\label{fig:branching}
\end{figure}

\begin{figure}[h]
\begin{center}
\includegraphics[angle=0,width=0.9\textwidth]{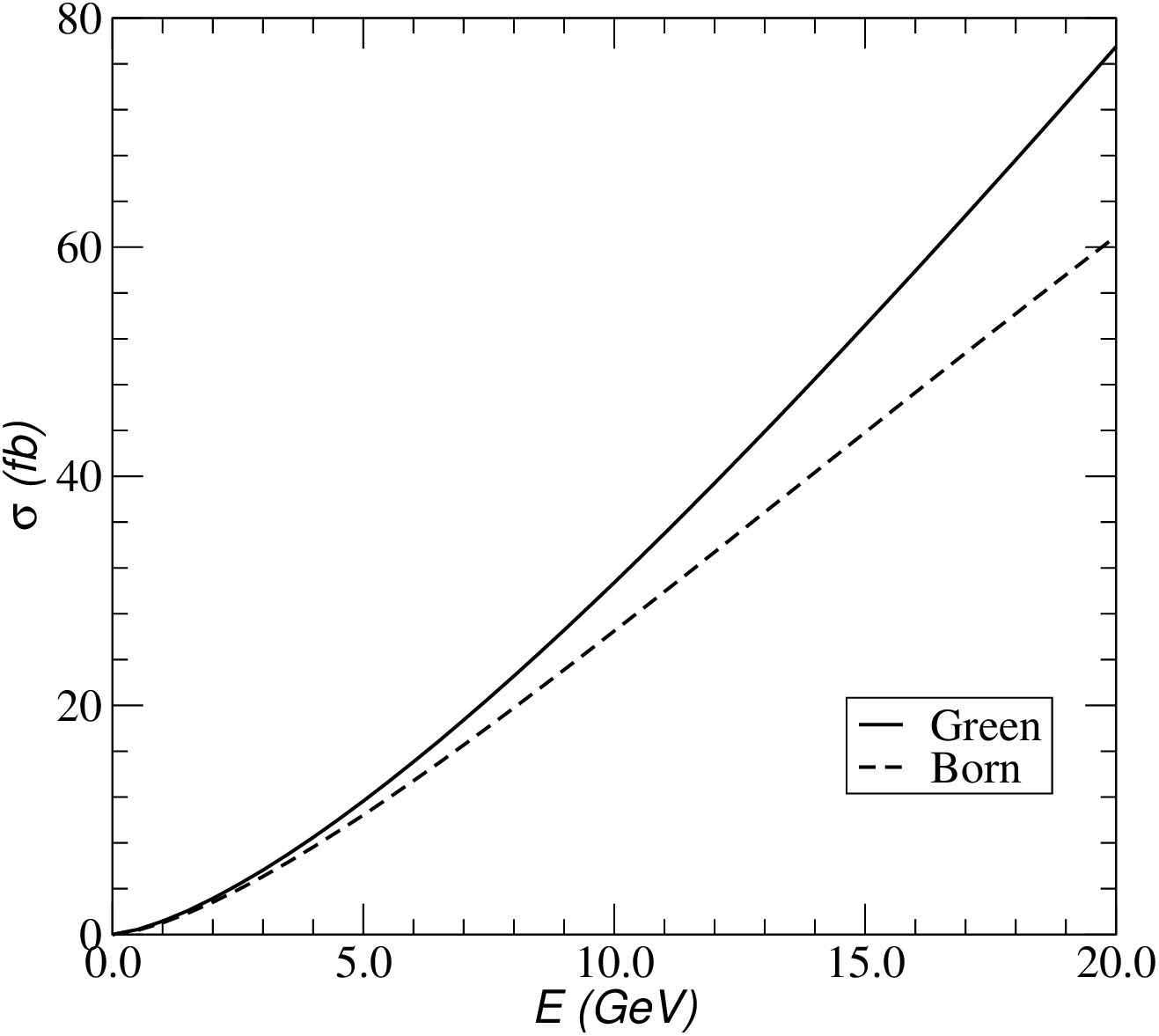}
\end{center}
\caption{ {Comparison of the Green function method and the Born level
expression for the cross section, for the right smuon. Here we assume that
$m_{{\smu}_R}=100 \gev$ and a displacement from threshold of up to $20 \gev$.}}
\label{fig:greenvsbornR}
\end{figure}

\begin{figure}[h]
\begin{center}
\includegraphics[angle=0,width=0.9\textwidth]{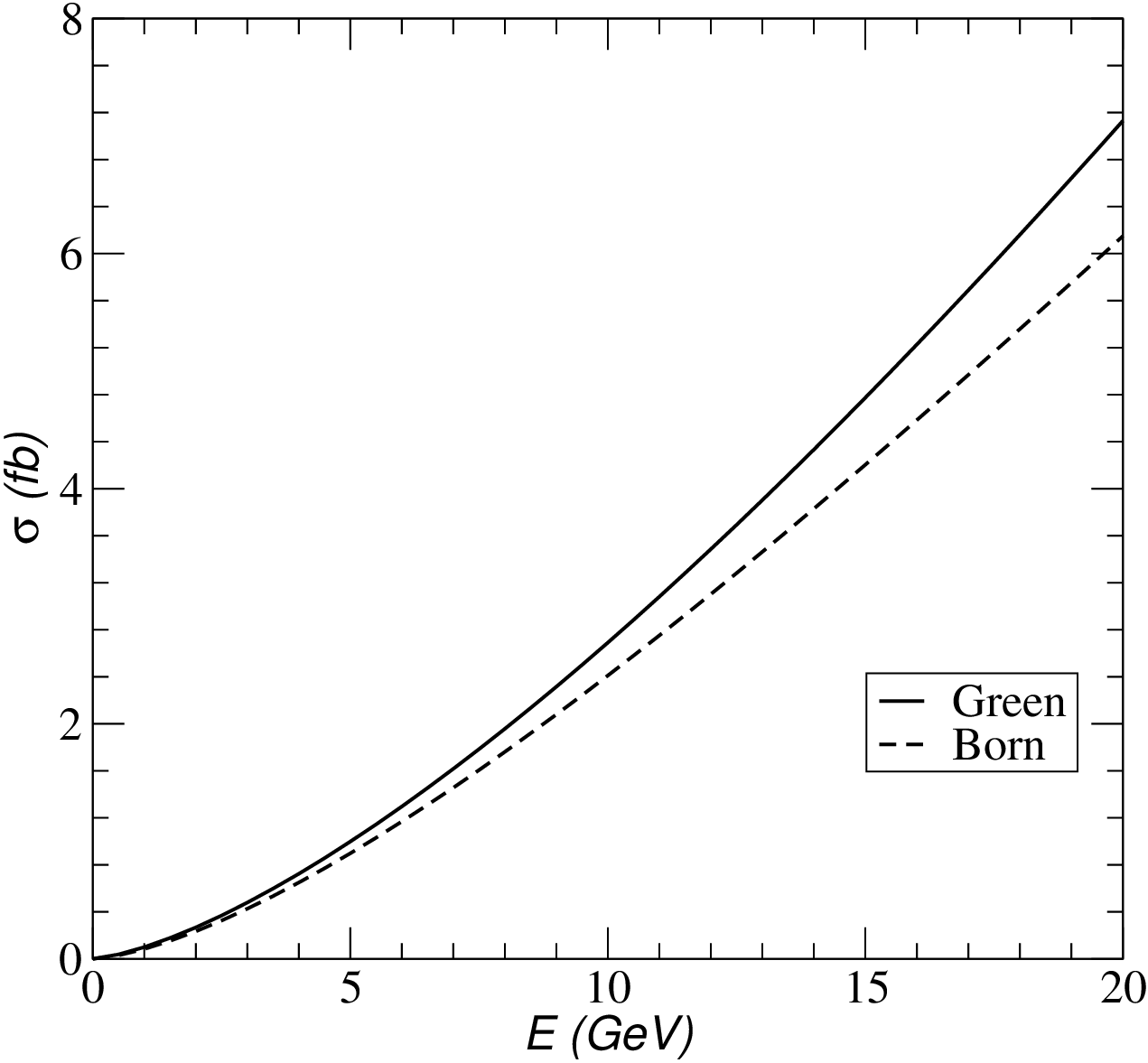}
\end{center}
\caption{ {Comparison of the Green function method and the Born level
expression for the cross section, for the right smuon. Here we assume that
$m_{{\smu}_R}=200 \gev$ and a displacement from threshold of up to $20 \gev$.}}
\label{fig:greenvsbornR200}
\end{figure}

\begin{figure}[h]
\begin{center}
\includegraphics[angle=0,width=0.9\textwidth]{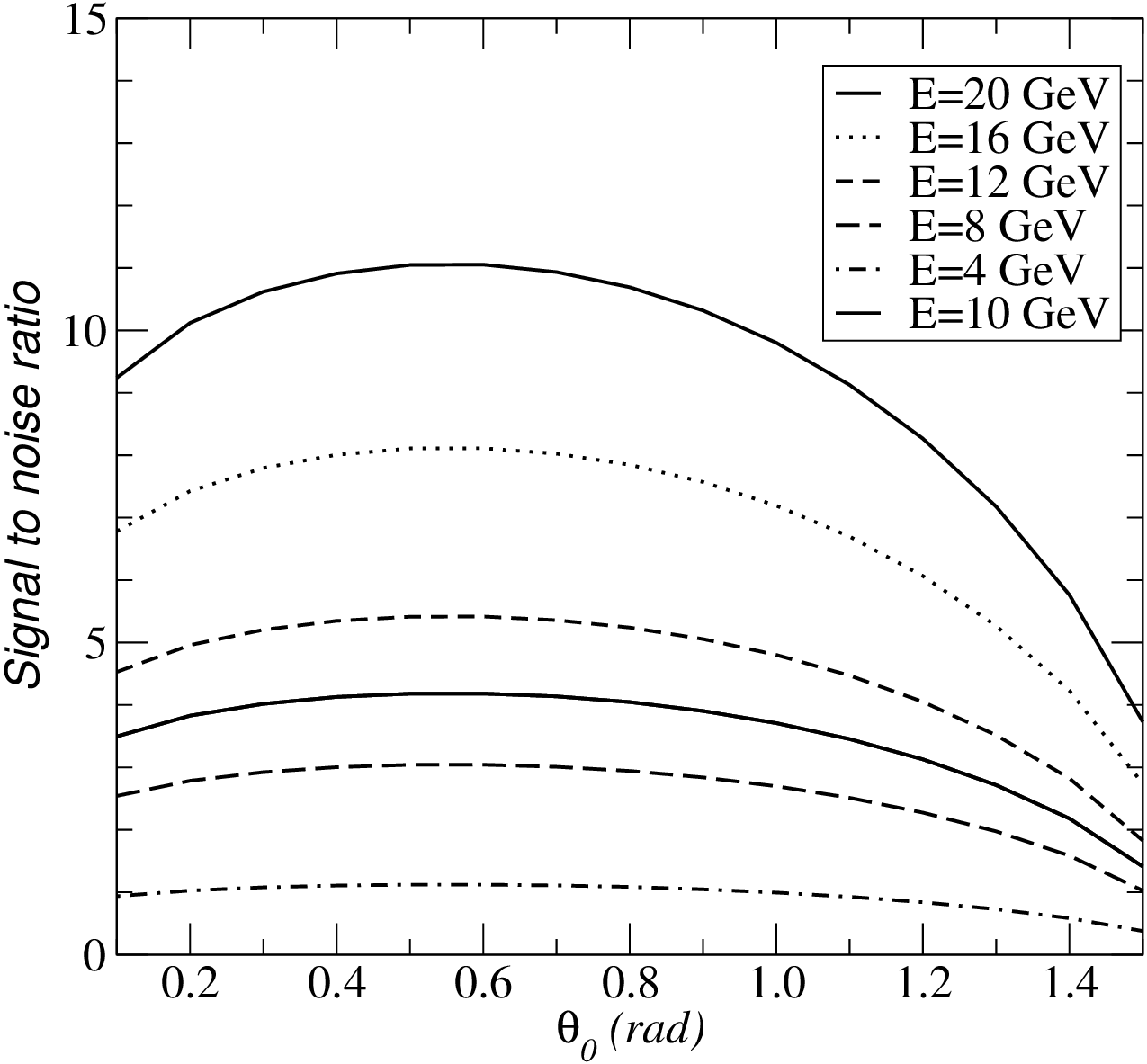}
\end{center}
\caption{
Statistical significance (signal to noise ratio) of the cross section
given by the Green function method and the $\gamma\gamma$ QED background 
as a function of
the angular cut angle $\theta_0$, for the right smuon $\smu_R$. We assume 
$m_{\smu_R}=100$ GeV; the various curves represent different displacements 
from threshold.}
\label{fig:greenvsnoiseR}
\end{figure}

\begin{figure}[h]
\begin{center}
\includegraphics[angle=0,width=0.9\textwidth]{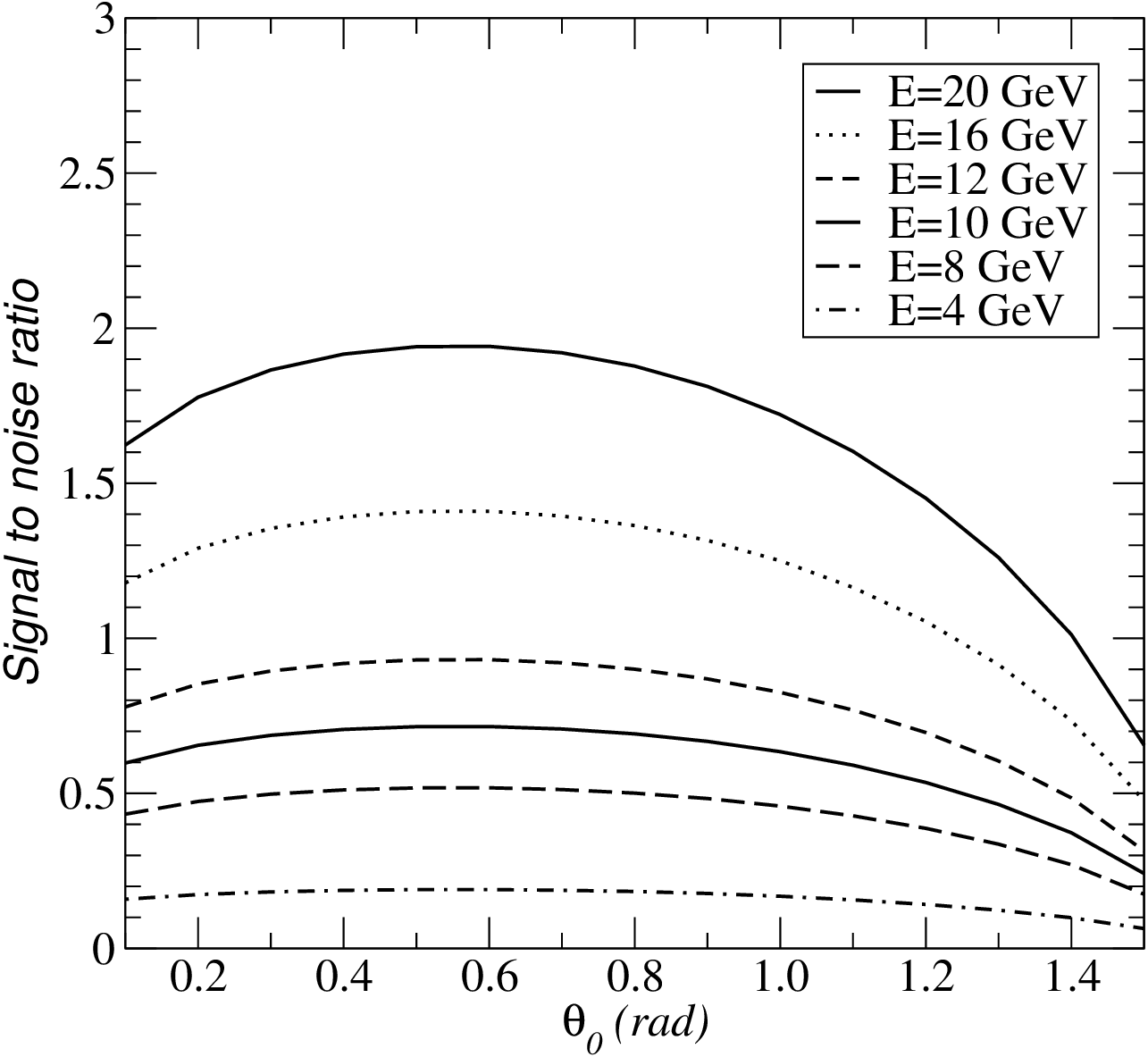}
\end{center}
\caption{
Statistical significance (signal to noise ratio) of the cross section
given by the Green function method and the $\gamma\gamma$ QED background 
as a function of
the angular cut angle $\theta_0$, for the right smuon $\smu_R$. We assume 
$m_{\smu_R}=200$ GeV; the various curves represent different displacements 
from threshold.}
\label{fig:greenvsnoiseR200}
\end{figure}

\begin{figure}[h]
\begin{center}
\includegraphics[angle=0,width=0.9\textwidth]{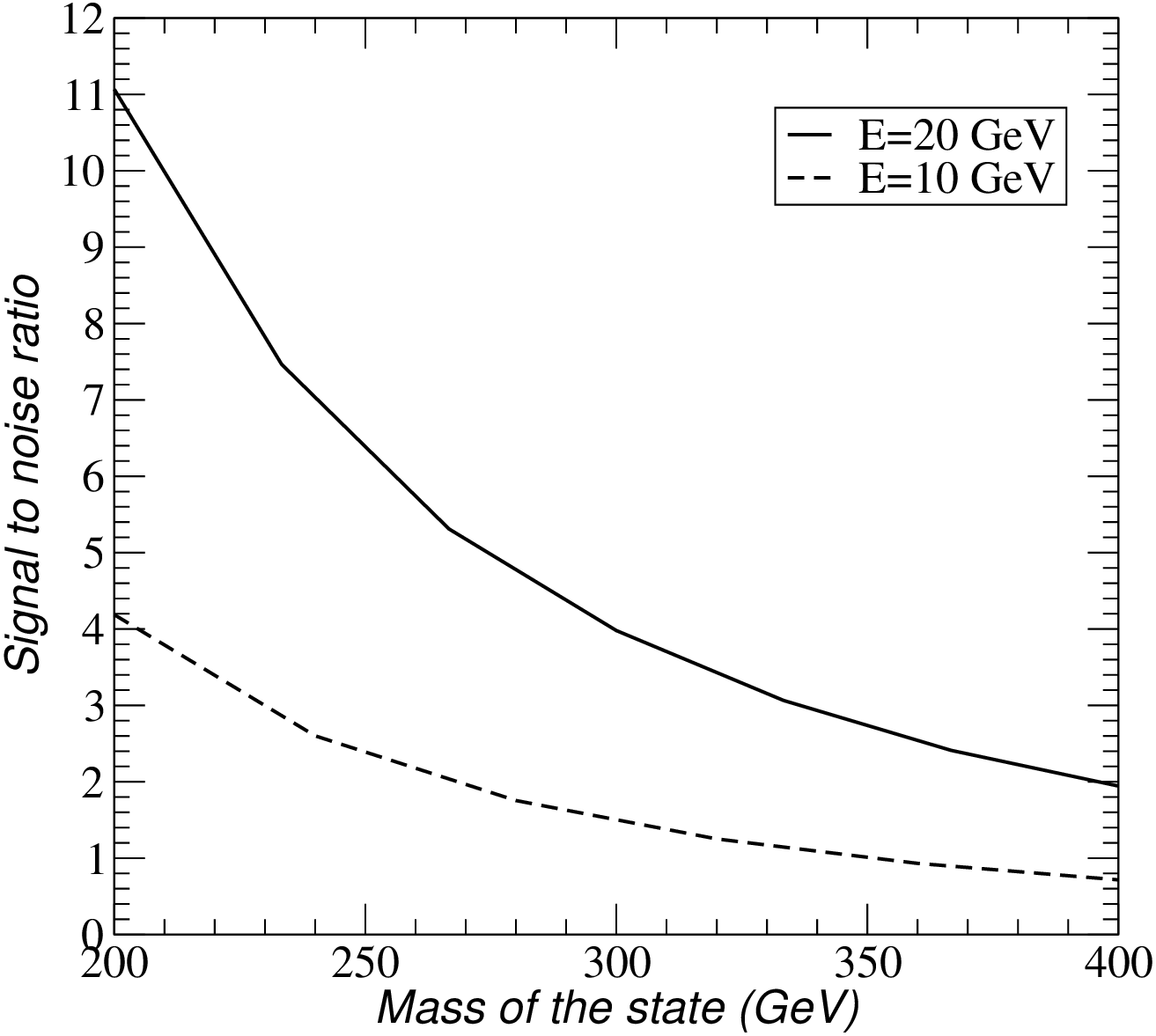}
\end{center}
\caption{
Statistical significance (signal to noise ratio) with respect to the mass of the bound state $M_B = 2 m_{NLSP}$. 
We show here the case of 
the right smuon $\smu_R$ bound state  produced at an energy $E=10$ and 
$E=20$ GeV above  threshold. The angular cut is assumed to be
at $\theta_0= 0.55$, maximal value of the signal to noise ratio.}
\label{fig6}
\end{figure}

\section{Conclusions}
\label{sec:conclu}
In this work a novel signature at the next LC is proposed for the detection 
of a sleptonium bound state (of a charged NLSP) assuming that the 
energy of the collider happens to be around the threshold.
It is well known that GMSB models are characterized by large regions of the
parameter space in which the NLSP is a charged slepton 
$\slep_R=(\widetilde{e}_R,\widetilde{\mu}_R,\widetilde{\tau}_1)$. 
While interesting signatures of the production of a pair of charged sleptons
NLSP have already been considered in the literature, 
here we discuss the formation, decay spectroscopy and possible 
detection of  the NLSP (charged slepton) 
{\em bound state} and its possible signature at the next LC.
For the sake of simplicity we have considered only the case of 
$\slep_R=\widetilde{\mu}_R$.
 
At an $e^+e^-$ collider the smuonium bound state
is produced in a 2P state. We assume a standard Coulombic interaction to 
describe the bound states and thereby estimate the energy levels and their 
separation which are needed to establish whether or not the formation 
takes place. 
As a criterion for formation we have chosen the more conservative condition, as opposed to the level gap criterion, obtained by requiring  that the 
revolution time be larger than the lifetime of the
rotating constituents.
We find out that the formation of the $2P$ bound state is assured for
$\sqrt{F_0} \ge 14.7 \, m_{\slep}$ and thus for all relevant values of
$\sqrt{F_0}$, the fundamental scale of SUSY breaking.
We have first analyzed the decay channels of the $2P$ state. These include:
a) decay into a pair of gravitinos, $2P \to \widetilde{G} \widetilde{G}$, 
b) decay to a pair of standard model fermions $2P \to f \bar{f}$ 
and  c) the dipole transition $2P \to 1S + \gamma $. Studying the partial 
widths and the branching ratios we find that for all 
practical purposes the $2P$ state decays to the $1S$ state emitting a photon 
whose energy is of a few MeV ($\Delta E_{2P-1S}$). 

In turn the decay channels of the $1S$ state are to the following final 
states: $\gamma \gamma, \gamma Z , ZZ, W^+W^-, hh$ and it turns out that 
the dominant decay is $1S \to \gamma\gamma$ as 
shown in Figs.~\ref{fig:partialwidths},\ref{fig:branching}.
Thus as the MeV photon from the $2P$ decay goes undetected because its energy is below the detector resolution, 
the following signal of the sleptonium bound state is 
defined \[ e^+e^- \to 2P 
\to 1S + \text{``soft $\gamma$''} \to \gamma \gamma\, , \]
where the observed final state is  {\em two hard back to back photons},
and ''practically'' no missing energy.

The bound state production cross section has been estimated using the 
Green function method. Due to the fact that there a bound state actually
exists, this effect accounts for a dramatically different cross section with
respect to the case of the Born production. For energies just under the
threshold, there are several peaks centred at the discrete energy levels of the
bound state. Approaching the threshold level from below, those peaks merge and
accumulate towards $E=0$ level. This fact translates to the energies above
threshold as well ($E>0$), increasing the cross section of the continuum. The
net effect is a larger cross section with respect to the naive Born
case. This effect grows with the strength of the coupling of the particles
which form the bound state itself. It is still very noticeable even for a weak
coupling like the sleptonium case, as it is larger than the Born 
cross section for about 10\% to 30\%,  the difference increasing with the 
displacement from threshold.

We have given an analysis of the two-photon signal comparing it to the QED 
two-photon cross-section (at leading order) 
defining a statistical significance depending on the 
angular cut which is introduced in order to reduce the QED background.
We find that at an energy offset of $E=20$ GeV from the threshold, 
the statistical significance $SS(\theta_0)$ 
(computed for an angular cut $\theta_0 =0.55$ which 
maximizes it)  goes from $SS=11$ at $m_{\slep_R} =100$ with 
$N_{sig} \approx 5000 $ to $SS=2 $ for  $m_{\slep_R} =200$ with 
$N_{sig} \approx 500 $.
 
Our study has been  done assuming a charged slepton co-NLSP scenario 
which is a peculiarity  of GMSB models. 
On the basis of the results discussed above we conclude 
that the study of a charged slepton NLSP {\em bound 
state } through the two-photon signature at the next LC has 
the potential to shed insights into the mechanism of SUSY breaking.\\
\mbox{} \\
{One of the authors (N.F.) wishes do dedicate this work to the memory
of Edmondo Pedretti}.

\end{document}